\documentclass[aps,prc,twocolumn,groupedaddress,showpacs]{revtex4-1}
\usepackage{graphicx}
\usepackage{comment}

\begin{document}

%\preprint{}

\title{Medium effects in proton-induced $K^{0}$ production at 3.5~GeV}



\author{G.~Agakishiev$^{7}$, O.~Arnold$^{10,9}$, D.~Belver$^{18}$, A.~Belyaev$^{7}$, 
J.C.~Berger-Chen$^{10,9}$, A.~Blanco$^{2}$, M.~B\"{o}hmer$^{10}$, J.~L.~Boyard$^{16}$, P.~Cabanelas$^{18}$, 
S.~Chernenko$^{7}$, A.~Dybczak$^{3}$, E.~Epple$^{10,9}$, L.~Fabbietti$^{10,9}$, O.~Fateev$^{7}$, 
P.~Finocchiaro$^{1}$, P.~Fonte$^{2,b}$, J.~Friese$^{10}$, I.~Fr\"{o}hlich$^{8}$, T.~Galatyuk$^{5,c}$, 
J.~A.~Garz\'{o}n$^{18}$, R.~Gernh\"{a}user$^{10}$, K.~G\"{o}bel$^{8}$, M.~Golubeva$^{13}$, D.~Gonz\'{a}lez-D\'{\i}az$^{5}$, 
F.~Guber$^{13}$, M.~Gumberidze$^{5,c}$, T.~Heinz$^{4}$, T.~Hennino$^{16}$, R.~Holzmann$^{4}$, 
A.~Ierusalimov$^{7}$, I.~Iori$^{12,e}$, A.~Ivashkin$^{13}$, M.~Jurkovic$^{10}$, B.~K\"{a}mpfer$^{6,d}$, 
T.~Karavicheva$^{13}$, I.~Koenig$^{4}$, W.~Koenig$^{4}$, B.~W.~Kolb$^{4}$, G.~Korcyl$^{3}$, 
G.~Kornakov$^{5}$, R.~Kotte$^{6}$, A.~Kr\'{a}sa$^{17}$, F.~Krizek$^{17}$, R.~Kr\"{u}cken$^{10}$, 
H.~Kuc$^{3,16}$, W.~K\"{u}hn$^{11}$, A.~Kugler$^{17}$, T.~Kunz$^{10}$, A.~Kurepin$^{13}$, 
V.~Ladygin$^{7}$, R.~Lalik$^{10,9}$, K.~Lapidus$^{10,9,\ast}$, A.~Lebedev$^{14}$, L.~Lopes$^{2}$, 
M.~Lorenz$^{8}$, L.~Maier$^{10}$, A.~Mangiarotti$^{2}$, J.~Markert$^{8}$, V.~Metag$^{11}$, 
J.~Michel$^{8}$, C.~M\"{u}ntz$^{8}$, R.~M\"{u}nzer$^{10,9}$, L.~Naumann$^{6}$, Y.~C.~Pachmayer$^{8}$, 
M.~Palka$^{3}$, Y.~Parpottas$^{15,f}$, V.~Pechenov$^{4}$, O.~Pechenova$^{8}$, J.~Pietraszko$^{4}$, 
W.~Przygoda$^{3}$, B.~Ramstein$^{16}$, A.~Reshetin$^{13}$, A.~Rustamov$^{8}$, A.~Sadovsky$^{13}$, 
P.~Salabura$^{3}$, A.~Schmah$^{a}$, E.~Schwab$^{4}$, J.~Siebenson$^{10,9}$, Yu.G.~Sobolev$^{17}$, 
B.~Spruck$^{11}$, H.~Str\"{o}bele$^{8}$, J.~Stroth$^{8,4}$, C.~Sturm$^{4}$, A.~Tarantola$^{8}$, 
K.~Teilab$^{8}$, P.~Tlusty$^{17}$, M.~Traxler$^{4}$, H.~Tsertos$^{15}$, T.~~Vasiliev$^{7}$, 
V.~Wagner$^{17}$, M.~Weber$^{10}$, C.~Wendisch$^{6,d}$, J.~W\"{u}stenfeld$^{6}$, S.~Yurevich$^{4}$, 
Y.~Zanevsky$^{7}$}

\affiliation{
(HADES collaboration) \\\mbox{$^{1}$Istituto Nazionale di Fisica Nucleare - Laboratori Nazionali del Sud, 95125~Catania, Italy}\\
\mbox{$^{2}$LIP-Laborat\'{o}rio de Instrumenta\c{c}\~{a}o e F\'{\i}sica Experimental de Part\'{\i}culas , 3004-516~Coimbra, Portugal}\\
\mbox{$^{3}$Smoluchowski Institute of Physics, Jagiellonian University of Cracow, 30-059~Krak\'{o}w, Poland}\\
\mbox{$^{4}$GSI Helmholtzzentrum f\"{u}r Schwerionenforschung GmbH, 64291~Darmstadt, Germany}\\
\mbox{$^{5}$Technische Universit\"{a}t Darmstadt, 64289~Darmstadt, Germany}\\
\mbox{$^{6}$Institut f\"{u}r Strahlenphysik, Helmholtz-Zentrum Dresden-Rossendorf, 01314~Dresden, Germany}\\
\mbox{$^{7}$Joint Institute of Nuclear Research, 141980~Dubna, Russia}\\
\mbox{$^{8}$Institut f\"{u}r Kernphysik, Goethe-Universit\"{a}t, 60438 ~Frankfurt, Germany}\\
\mbox{$^{9}$Excellence Cluster 'Origin and Structure of the Universe' , 85748~Garching, Germany}\\
\mbox{$^{10}$Physik Department E12, Technische Universit\"{a}t M\"{u}nchen, 85748~Garching, Germany}\\
\mbox{$^{11}$II.Physikalisches Institut, Justus Liebig Universit\"{a}t Giessen, 35392~Giessen, Germany}\\
\mbox{$^{12}$Istituto Nazionale di Fisica Nucleare, Sezione di Milano, 20133~Milano, Italy}\\
\mbox{$^{13}$Institute for Nuclear Research, Russian Academy of Science, 117312~Moscow, Russia}\\
\mbox{$^{14}$Institute of Theoretical and Experimental Physics, 117218~Moscow, Russia}\\
\mbox{$^{15}$Department of Physics, University of Cyprus, 1678~Nicosia, Cyprus}\\
\mbox{$^{16}$Institut de Physique Nucl\'{e}aire (UMR 8608), CNRS/IN2P3 - Universit\'{e} Paris Sud, F-91406~Orsay Cedex, France}\\
\mbox{$^{17}$Nuclear Physics Institute, Academy of Sciences of Czech Republic, 25068~Rez, Czech Republic}\\
\mbox{$^{18}$LabCAF. F. F\'{\i}sica, Univ. de Santiago de Compostela, 15706~Santiago de Compostela, Spain}\\ 
\\
\mbox{$^{a}$ also at Lawrence Berkeley National Laboratory, ~Berkeley, USA}\\
\mbox{$^{b}$ also at ISEC Coimbra, ~Coimbra, Portugal}\\
\mbox{$^{c}$ also at ExtreMe Matter Institute EMMI, 64291~Darmstadt, Germany}\\
\mbox{$^{d}$ also at Technische Universit\"{a}t Dresden, 01062~Dresden, Germany}\\
\mbox{$^{e}$ also at Dipartimento di Fisica, Universit\`{a} di Milano, 20133~Milano, Italy}\\
\mbox{$^{f}$ also at Frederick University, 1036~Nicosia, Cyprus}\\
}

\author{
%\vspace{5mm}
T. Gaitanos$^{19}$, J. Weil$^{20}$\\
%\vspace{20mm}
}
\affiliation{
\mbox{$^{19}$ Institut f\"ur Theoretische Physik I, 35392~Giessen, Germany}\\
\mbox{$^{20}$ Frankfurt Institute for Advanced Studies, 60438~Frankfurt am Main, Germany}\\
\\
\mbox{$^{\ast}$ corresponding author: kirill.lapidus@ph.tum.de}
}


\date{\today}

\begin{abstract}
We present the analysis of the inclusive $K^{0}$ production in p+p and p+Nb collisions measured with the HADES detector at a beam kinetic energy of 3.5 GeV. Data are compared to the GiBUU transport model. The data suggest the presence of a repulsive momentum-dependent kaon potential as predicted by the Chiral Perturbation Theory (ChPT). For the kaon at rest and at normal nuclear density, the ChPT potential amounts to $\approx 35$~MeV. A detailed tuning of the kaon production cross sections implemented in the model has been carried out to reproduce the experimental data measured in p+p collisions. The uncertainties in the parameters of the model were examined with respect to the sensitivity of the experimental results from p+Nb collisions to the in-medium kaon potential.
\end{abstract}

%\keywords{}

\maketitle

%\vspace{5mm}

\section{Introduction}
Properties of hadrons immersed in a strongly interacting environment were the subject of intense theoretical and experimental studies over the last decades \cite{Tolos:2013qv}. At non-zero baryonic densities gradual restoration of the spontaneously broken chiral symmetry is expected \cite{Klimt:1990ws}, characterized by the melting of the quark condensate and leading to the modification of hadron spectral functions. Measurements of light vector mesons in the nuclear medium are a well known example of searches in this direction \cite{Leupold:2009kz}. In the pseudoscalar sector a particular attention is attracted to (anti)kaons that appear as the Goldstone bosons of spontaneously broken chiral symmetry \cite{Fuchs:2005zg}.

A large number of experiments searched for nuclear matter effects in collisions of heavy ions, where baryonic densities $\rho_{B}$ considerably exceeding normal nuclear density can be achieved \cite{Klimt:1990ws}. However, already at normal nuclear density $\rho_{0}$, that can be probed in proton-nucleus collisions, effects of the nuclear environment are expected to take place \cite{Post:2003hu}. An advantage of such an approach is a fixed density profile of the target nucleus that does not evolve in the course of the collision, contrary to the case of heavy-ion collisions.

The kaons ($K^{+}, K^{0}$) are peculiar probes of the nuclear matter effects: since they contain an anti-strange valence quark, they do not form baryon resonances when interacting with nucleons and propagate in nuclear matter relatively freely with a mean free path of $\lambda\approx5$~fm for kaon momenta $p_K<900$~MeV$/c$. The low-density theorem relates the kaon self energy in a nuclear environment with free kaon-nucleon scattering amplitudes. Application of the low density theorem yields a repulsive kaon-nucleus potential of $\sim$20--30~MeV at normal nuclear density for kaons at rest. (These values come from earlier estimates, e.~g. in \cite{Kolomeitsev:1995xz}.)
A theoretical calculation performed in \cite{Korpa:2004ae} gives, within the low-density theorem, a potential of 25~MeV and a larger value of 35~MeV when using a more involved self-consistent approach. Another important theoretical finding is that the kaon spectral function remains narrow at moderate nuclear densities, making the in-medium kaon a well-defined quasi-particle that can be propagated in the transport-model approach.

Several recent experiments addressed the issue of the kaon-nucleus potential. Kaon production on different nuclear targets in pion- and proton-induced reactions was studied by the FOPI \cite{Benabderrahmane:2008qs} ($\pi^{-} A$) and ANKE \cite{Buescher:2004vn} ($pA$) collaborations. From comparisons with transport-model simulations an average repulsive potential of 20$\pm$5 MeV was inferred. The HADES collaboration reported on a measurement of $K^{0}_{S}$ in a medium-sized colliding system, Ar+KCl, at a beam energy of 1.76 GeV/nucleon \cite{Agakishiev:2010zw} (see also \cite{Agakishiev:2009ar} for the $K^{+}$ data). An average potential value of $39^{+8}_{-2}$~MeV was inferred from these data. These deviating values (both quoted for $\rho_{B} = \rho_{0}$ and $p_{K} = 0$) call for a further experimental effort. We note that both analyses employed a simple parameterization of the potential; this issue is discussed further in Section \ref{sec:pNb}.

For results obtained by the KaoS and the FOPI collaborations in analyses of the azimuthal anisotropy of the (anti)kaon emission and its sensitivity to the in-medium potentials we refer to \cite{Shin:1998hm, Uhlig:2004ue, Crochet:2000fz}.

The kaon, produced inside a nucleus, is not influenced solely by a mean-field potential, but also undergoes scattering on individual nucleons. As mentioned above, for $p_{K} < 900$~MeV/c the kaon-nucleon interaction is rather weak ($\sigma_{KN} \approx 12.5$~mb) and is limited to the elastic scattering $KN \to KN$, including the charge-exchange process, and the first inelastic channel $KN \to KN\pi$. 
Transport models are incorporating free kaon-nucleon cross sections, 
which are known well in a broad energy range.
However, there are claims for an in-medium modification of the kaon-nucleon cross sections: experiments with a kaon beam interacting with nuclear targets revealed that the usage of the free kaon-nucleon scattering amplitudes leads to an underestimation of the $K^{+}$-nucleus reaction cross sections at the level of 15-20\% \cite{Friedman:2007zza}. A number of mechanisms was proposed to explain this difference: swelling of nucleons in nuclear matter \cite{Siegel:1985hp}, in-medium modification of exchanged vector mesons \cite{Brown:1988gu}, contribution of meson exchange-current diagrams \cite{Jiang:1992wn}, formation of a pentaquark state \cite{Gal:2004cx, Tolos:2005jg}, etc.

As discussed in \cite{Sood:2011qj}, both the kaon-nucleus potential and the in-medium kaon-nucleon scattering affect the final state phase-space distributions of kaons. An ambiguity in the strengths of these two contributions hampers an interpretation of results obtained in heavy-ion collisions. Thus, new experimental data that allow to constrain the two effects is of high importance. 
In our study of the $K^{0}$ production on a Nb target bombarded by a proton beam with a kinetic beam energy of 3.5 GeV we address the issue of the in-medium potential and check whether the usage of the free KN interaction cross sections allows for a good description of the experimental data.
As a reference system, which allows to constrain production cross sections, we use the inclusive measurement of $K^{0}$'s in p+p collisions at the same beam energy.

Together with an analysis of the exclusive $K^{0}$ production in proton-proton reactions \cite{Agakishiev:2013yyy}, the present data provide a baseline for measurements of strangeness production in nuclear collisions at beam energies of 2-8~GeV/nucleon available at the future FAIR facility.

The paper is organized as follows. Section~\ref{sec:exp} gives a brief information about the detector system. The analysis procedure is described in Section~\ref{sec:dat_an}. Sections~\ref{sec:pp} and \ref{sec:pNb} contain experimental results and their interpretation within the Giessen Boltzmann-Uehling-Uhlenbeck (GiBUU) transport model \cite{Buss:2011mx}. Section~\ref{sec:sum} concludes the paper with a summary of main findings.

\section{the Experiment \label{sec:exp}}

Data for the analysis were collected with the {\bf H}igh-{\bf A}cceptance {\bf D}i-{\bf E}lectron {\bf S}pectrometer (HADES). It is a versatile charged-particle detector currently operating at the SIS18 synchrotron (GSI Helmholtzzentrum, Darmstadt) in the region of beam kinetic energies of 1--2~GeV/nucleon for nucleus-nucleus collisions, up to 3.5~GeV in proton-induced reactions. The detector covers polar angles from 18$^{\circ}$ to 85$^{\circ}$ degrees and a large portion of the azimuthal angle; the momentum resolution of the spectrometer is $ \Delta p/p\approx3\%$. The main components of the experimental setup are a superconducting magnet, four planes of Multiwire Drift Chambers (MDC) used for the tracking of charged particles, a Time-of-Flight wall and a hadron-blind Ring Imaging Cherenkov (RICH) detector; a detailed description of the detector ensemble can be found in \cite{Agakishiev:2009am}.

In 2007 a measurement of proton-proton collisions at a kinetic beam energy of 3.5 GeV was performed: the beam with an average intensity $\sim 1\times10^{7}$ particles/s was incident on a liquid hydrogen target with a density of 0.35~g$/$cm$^{2}$ and a total interaction probability of $\sim 0.7 \%$. In total, $1.2 \times 10^{9}$ events were collected. 
In 2008 the proton beam with the same characteristics was directed onto a segmented $^{93}$Nb target. 
Overall, $4 \times 10^{9}$ events had been taken. In both runs, the first-level trigger (LVL1) required at least three hits (M3) in the Time-of-Flight wall.

\section{Data Analysis \label{sec:dat_an}}
\subsection{Kaon identification}
\begin{figure}[h]
\includegraphics[width=0.5\textwidth]{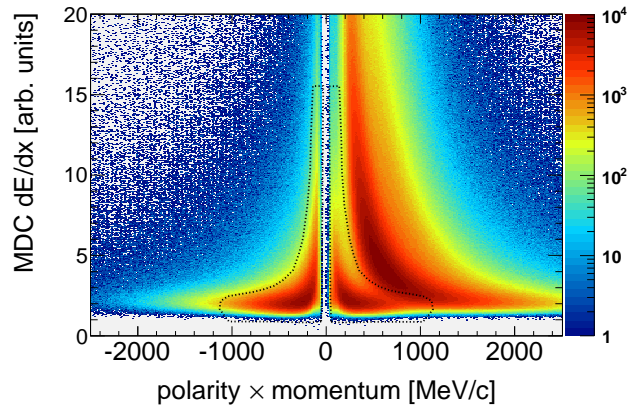}
\caption{\label{fig:pNb_dEdx} (Color online) Total specific energy loss in the MDCs versus momentum of the particle. Charged pions are selected with the help of two-dimensional dE/dx cuts (shown by dotted curves).}
\end{figure}

The $K^{0}$ is identified by its short-lived component (50\%) $K^{0}_{S}$ ($c\tau\simeq2.68$~cm) that decays weakly into a $\pi^{+}\pi^{-}$ pair with a branching ratio of 69.2\%. Charged pions are identified with the help of two-dimensional cuts in the $(dE/dx)_{MDC}$ vs momentum plane, where $(dE/dx)_{MDC}$ is the measured particle's energy loss in the Multiwire Drift Chambers (Fig.~\ref{fig:pNb_dEdx}). This is the only information used for particle identification in this analysis. All intersections and points of closest approach of oppositely charged pion tracks are considered as candidates for secondary vertices where the $K^{0}_S$ decays into a pair of charged pions. In order to suppress the combinatorial background the following topological cuts are applied: i)~a cut on the distance between the primary to the secondary vertex $d(K^{0}_S - V)>25$~mm, ii)~a cut on the distance of closest approach between two pion tracks $d_{\pi^{+}-\pi^{-}}<7$~mm, and iii)~a cut on the distance of closest approach between either of the extrapolated pion tracks and the primary vertex $DCA_{\pi}>7$~mm. The values of the applied cuts were optimized in order to provide the best compromise between the $K^{0}_S$ signal strength and the signal-to-background ratio.

\begin{figure}[htb]
\includegraphics[width=0.4\textwidth]{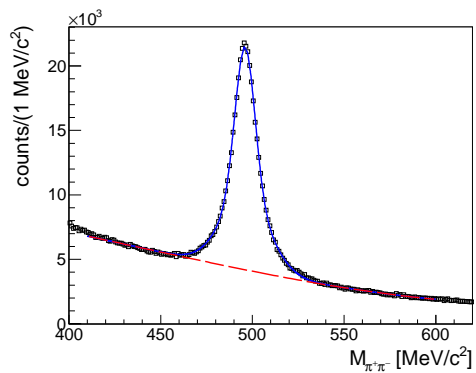}
\caption{\label{fig:pNb_signal} (Color online) Invariant mass distribution of $\pi^{+}\pi^{-}$ pairs in p+Nb collisions after topological cuts showing a $K^{0}_{S}$ signal, characterised by a (extracted by a fit) mass of $M_{K^0_S} = 495.2 \pm 0.02$~MeV/c$^{2}$ and a width of $\sigma_{K^0_S} = 7.29 \pm 0.05$~MeV/c$^{2}$. The number of reconstructed $K^0_S$ amounts to $N_{K^{0}_{S}} = \left(325.0 \pm 0.7\right) \times 10^{3}$.}
\end{figure}

The resulting $\pi^{+}\pi^{-}$ invariant mass spectrum for p+Nb collisions is presented in Fig.~\ref{fig:pNb_signal}. A clear high-statistic signal corresponding to the $K^{0}_{S}$ is visible on top of a combinatorial background. In order to extract the $K^{0}_{S}$ signal, a simultaneous two-component (signal + background) fit is performed, which models the background as a sum of a Landau and a polynomial function, and the signal as a sum of two Gaussian functions.

The 3-dimensional kaon phase space can be fully described by employing, for example, the following variables: transverse momentum $p_{t}$, rapidity $y$ and azimuthal angle $\varphi$. The further analysis is performed in a 2-dimensional phase space; all distributions are integrated over the azimuthal angle, since both colliding systems (p+p and p+Nb) possess rotational symmetry with respect to the beam axis (neither target nor beam is polarised).

\subsection{Efficiency correction}

Extracted doubly differential yields of the $K^{0}_{S}$ have to be corrected for the finite efficiency of the 
analysis procedure that includes track reconstruction, pion identification and topological cuts. For this purpose, dedicated simulations have been carried out. For the p+p case a cocktail of the $K^{0}$ production channels based on the available world data was simulated with the Monte Carlo event generator Pluto \cite{Frohlich:2007bi}. In the next step, a high statistics sample of simulated data was used as input for the {\sc geant3} package, which models the response of the experimental apparatus. Finally, the simulated events went through the very same analysis chain as the experimental ones. This procedure allowed to calculate losses due to finite acceptance and efficiency and correct for them. The efficiency in a particular element of the kaon phase space (including the acceptance losses) is given by the ratio \begin{equation}
\label{eq:eff_acc}
\epsilon\left(p_{t},y\right) = \frac{f_{out}\left(p_{t},y\right)}{f_{in}\left(p_{t},y\right)},
\end{equation}  
where $f_{in}\left(p_{t},y\right)$ is the generated and $f_{out}\left(p_{t},y\right)$ is the reconstructed phase space population of kaons.

A typical reconstruction efficiency results in $\epsilon\left(p_{t}, y\right) = 4-8$\,\%. It was checked that the resulting efficiency correction matrix is not sensitive to the changes in strengths of different contributing reactions. Note, moreover, that the acceptance losses are corrected for only in the region of the phase space where experimental data are available, i.e. no extrapolation into the unmeasured kinematical region takes place at this stage.

The case of p+Nb reactions was treated analogously, but the UrQMD transport code \cite{Bass:1998ca} was used to generate events containing kaons and a background of charged particles.

\subsection{Absolute normalization}
Differential kaon yields measured in p+p reactions are normalized absolutely using the factor obtained from the analysis of the elastic p+p scattering in the HADES acceptance \cite{HADES:2011ab}. For the p+Nb case the total reaction cross section $\sigma_{pNb} = 848 \pm 127$~mb is provided by the analysis of the charged pion production \cite{Tlusty2010, Agakishiev:2012vj}.  

The first-level multiplicity trigger M3
slightly enhances the average centrality of p+Nb collisions and introduces a bias on the measured $K^{0}_{S}$ multiplicity. This bias has been evaluated using the UrQMD transport model \cite{Bass:1998ca} and the data were corrected. Thus, the presented p+Nb measurements correspond to all inelastic collisions.

\section{Elementary proton-proton reactions \label{sec:pp}}
\begin{figure*}[t]

\includegraphics[width=0.85\textwidth]{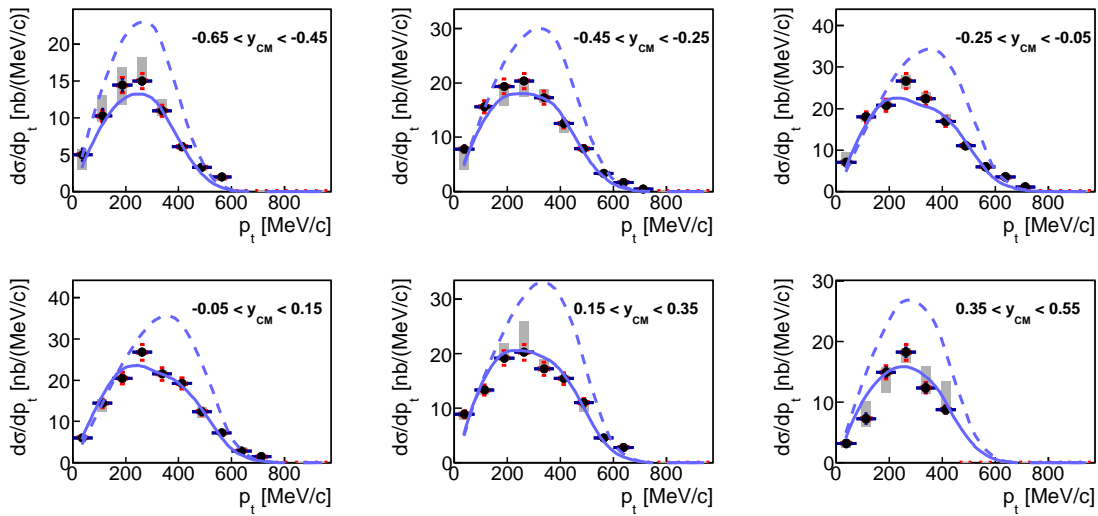} %new 0.67 0.4 0.7 0.84 

\caption{\label{fig:pp_pt_gibuu} (Color online) $K^{0}_S$ transverse momentum spectra in p+p collisions: experimental data (black circles) and GiBUU transport model simulations (dashed curves --- original resonance model \cite{Tsushima:1998jz}, solid curves --- modified resonance model, see text). Systematic uncertainties due to the efficiency correction and absolute normalization are indicated by shaded boxes and red horizontal dashes, respectively.}

\end{figure*}

Efficiency-corrected and absolutely normalized transverse momentum distributions of $K^{0}_{S}$'s reconstructed in p+p collisions in six rapidity bins ($y_{CM}\in(-0.65, 0.55)$, $\Delta y_{CM} = 0.2$), where $y_{CM}$ is the rapidity in the nucleon-nucleon center of mass reference frame, are shown in Fig.~\ref{fig:pp_pt_gibuu}. The systematic uncertainties due to the acceptance and efficiency correction procedure were evaluated by varying by $\pm 20\%$ all topological cuts described above. They are shown as shaded boxes on Fig.~\ref{fig:pp_pt_gibuu}. Larger systematic uncertainties are typical for the data points with low statistics due to the worse stability of the signal extraction by a fit. Another source of systematic uncertainties is the absolute normalization to the number of elastic proton-proton collisions; these uncertainties are indicated in Fig.~\ref{fig:pp_pt_gibuu} by red dashes.

The GiBUU transport model (version 1.6) \cite{Buss:2011mx} simulations are shown on Fig.~\ref{fig:pp_pt_gibuu} as well. In this model kaon production in baryon-baryon collisions is treated either in the resonance model, developed in \cite{Tsushima:1998jz}, or in the {\sc pythia} string fragmentation framework; details can be found in \cite{Buss:2011mx}. 

The strangeness sector of the GiBUU resonance model is given by the Tsushima model \cite{Tsushima:1998jz}, 
in which the elementary kaon production channels $NN \to BYK$
(with $B = N, \Delta$ and $Y = \Lambda, \Sigma$) are assumed to proceed via the formation and
decay of various $N^{*}$ and $\Delta^{*}$ resonances. The corresponding cross sections are calculated in an effective-Lagrangian approach. The GiBUU implementation uses the final cross-section parameterizations from \cite{Tsushima:1998jz} and does not explicitly produce and propagate the intermediate resonances; a 3-body ($BYK$) production always takes place. It should be noted that the non-strange part of the GiBUU resonance model \cite{Buss:2011mx, Weil:2012ji} uses a different set of resonances than considered in \cite{Tsushima:1998jz} (without strange
decay modes). 
Therefore, the strangeness sector of the resonance model is disconnected from the non-strange part.

Calculations of the kaon production in proton-proton collisions based on the resonance model are shown on Fig.~\ref{fig:pp_pt_gibuu} by the dashed curves. Apparently, the resonance model significantly overestimates the inclusive production of $K^{0}$ mesons. This observation is consistent with the fact that the strengths of a number of exclusive kaon production channels are overestimated in this model \cite{Tsushima:1998jz}.

First of all, the contribution of the channel $p + p \to p + \pi^{+} + \Lambda + K^{0}$ is strongly overestimated (in the resonance model it goes exclusively through the reaction $p + p \to \Delta^{++}(1232) + \Lambda + K^{0}$) both at beam energies lower \cite{Nekipelov:2006if} and higher \cite{Bierman:1966zz, Alexander:1967zz, Klein:1970ri} than studied in our analysis. In order to fit our inclusive spectra and, at the same time be consistent with reported measurements at other energies, the cross section of this channel and the channel $p + p \to \Delta^{++}(1232) + \Sigma^{0} + K^{0}$ should be scaled by factors of 0.42 and 0.72, respectively. Similarly, the parameterizations given by the resonance model overestimate the data for the 3-body channel $p + p \to \Sigma^{+} + p + K^{0}$ in the region $\sqrt{s}=2.57-2.83$~GeV \cite{AbdelBary:2012vw} by a factor of $\approx1.5$. The cross section of this channel has been scaled down accordingly. It should be stressed that the adjusted cross sections are consistent with the results of the exclusive analysis of kaon production in the 4-body channels $p + p \to p + \pi^{+} + Y + K^{0}$ \cite{Agakishiev:2013yyy}. 

For all 3-body reactions $N + N \to N + Y + K$ we use a modified phase-space distribution following suggestions made in \cite{Larionov:2005eb, Zheng:2002mj, Li:1997zb}. These modifications account for the observed angular anisotropy of kaons in the $NYK$ final state and soften their momentum spectra.

At the beam energy of 3.5~GeV, corresponding to $\sqrt{s} = 3.18$~GeV, phase space opens for the kaon production channels with 5-body final states. For example, this energy is well above the threshold for the channel $p + p \to p + \pi^{+} + \Sigma^{+} + \pi^{-} + K^{0}$ with $\sqrt{s_{thr.}} \approx 2.9$~GeV. Such channels with two pions in the final state are not considered in the original resonance model~\cite{Tsushima:1998jz}. In order to include the contribution by the 5-body channels, the reactions $p + p \to \Delta(1232) + Y^{*} + K^{0}$ have been added, where $Y^{*}$ stays for the $\Sigma(1385)$ or the $\Lambda(1405)$ hyperon, which decay mainly in $\Lambda \pi$ and $\Sigma \pi$ pairs, respectively. Cross sections for these channels were tuned such as to reproduce the low-$p_{t}$ component of the measured inclusive spectra; an additional constraint was set that the cross sections for these channels should be lower than those for the $p + p \to N + Y^{*} + K$ reactions, reported in \cite{Agakishiev:2011qw, Agakishiev:2012qx, Agakishiev:2012xk}. The latter reactions are not included in the model explicitly, since they result in a 4-body final state already populated by reactions $p + p \to \Delta(1232) + Y + K$. All modifications of the production cross sections are summarized in Table~\ref{table:tableK1}. The exact parameterization for the channels $p + p \to \Delta(1232) + Y^{*} + K^{0}$ as a function of the nucleon-nucleon collision energy $s$ is given by the following formula:
\begin{equation}
\sigma(p + p \to \Delta + Y^{*} + K^{0}) = a \left(\frac{s}{s_0} - 1 \right)^{b} \left( \frac{s_{0}}{s} \right)^{c},
\end{equation}
where $a = \left(8.5, 3.1\right)$ [mb], $b = (2.842, 2.874)$, $c = (1.960, 2.543)$, $s_{0} = (9.356, 8.889)$ [GeV$^{2}$] for the $\Lambda(1405)$ and $\Sigma(1385)$ channel, respectively.

\begin{table}[b]%The best place to locate the table environment is directly after its first reference in text
\caption{\label{table:tableK1}%
Cross sections for $K^{0}$ production channels in p+p collisions at $E_{beam}^{kin.} = 3.5$~GeV. All values are in $\mu$b. The numbers in brackets are scaling factors that were applied to the values given by the resonance model \cite{Tsushima:1998jz} (Tsushima \emph{et al.}). The last three reactions channels are not considered in the original model.}
\begin{ruledtabular}
\begin{tabular}{lll}
Reaction, $p + p\to$ & Tsushima & Present work \\
\colrule
$p+ \Sigma^{+} + K^{0}$ & 37.8 & 26.5 (0.70)\\
\colrule
$p + \pi^{+} + \Lambda + K^{0}$ & 75.9 & 31.9 (0.42)\\
$p  + \pi^{+} + \Sigma^{0} + K^{0}$ & 24.6 & 17.7 (0.72)\\
$p  + \pi^{0} + \Sigma^{+} + K^{0}$ & 10.9 & 7.8 (0.72)\\
$n  + \pi^{+} + \Sigma^{+} + K^{0}$ & 5.5 & 3.9 (0.72)\\
\colrule
$\Delta^{++} + \Lambda(1405) + K^{0}$ & n/a & 5.3\\
$\Delta^{++} + \Sigma(1385)^{0} + K^{0}$ & n/a & 3.5\\
$\Delta^{+} + \Sigma(1385)^{+} + K^{0}$ & n/a & 2.3\\
\end{tabular}
\end{ruledtabular}
\end{table}

Calculations with the modified resonance model are shown in Fig.~\ref{fig:pp_pt_gibuu} (solid lines). The modified model describes the experimental spectra well.
Thus, the kaon production in proton-proton collisions is fixed in the model by tuning the corresponding cross sections such that they reproduce our measurement.

\begin{figure}[h]
\includegraphics[height=0.35\textwidth]{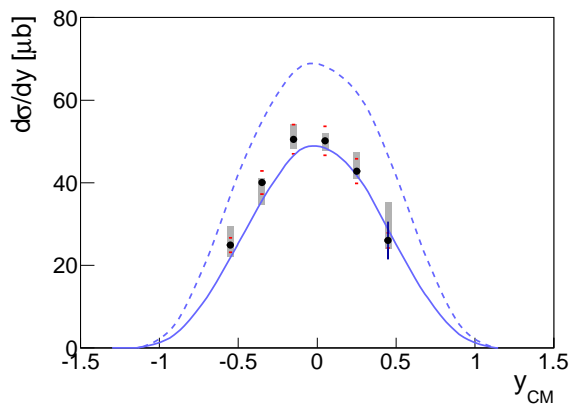}
\caption{\label{fig:pp_y_Pluto} (Color online) $K^{0}_S$ rapidity distribution in p+p collisions (black circles) and GiBUU transport model simulations (dashed curve --- original resonance model \cite{Tsushima:1998jz}, solid curve --- modified resonance model, see text).}
\end{figure}

The measured transverse momentum spectra allow the reconstruction of the rapidity density in the covered phase space.
The integrated yield per rapidity bin is calculated in the following way: the sum of all measured data points is taken and for the extrapolation to the unmeasured region of high $p_{t}$-values a Boltzmann fit is used. The resulting $dN/dy$ spectrum is shown in Fig.~\ref{fig:pp_y_Pluto} together with the calculations according to the original (dashed line) and modified (solid line) resonance models. The modified version of the resonance model reproduces very well both the shape of the rapidity distribution and the total yield of kaons. The rapidity distribution is symmetric with respect to mid-rapidity $y_{CM} = 0$: a fit with a Gaussian function delivers mean value of $M = -0.03$ and $\chi^{2}/NDF = 4.3/3$; this is an important check of the validity of the acceptance and efficiency corrections applied to the data.

Summation over the measured data points in the rapidity spectrum, extrapolation into the unmeasured region with a Gaussian fit and multiplication by a factor of 2 (for taking into account the 50\% branching of $K^{0}$ into $K^{0}_S$) delivers the total cross section of the inclusive $K^{0}$ production in proton-proton collisions:
\begin{equation}
\sigma(p+p \to K^{0}+X ) = 112.7 \pm 2.3 _{-1.0}^{+4.5} \pm 7.9~\mu \text{b}.
\end{equation}
The first quoted uncertainty has statistical origin, the second (asymmetric) is the systematic uncertainty originating from the variation of topological cuts, and the third one is the systematic uncertainty stemming from the absolute normalization.
An assumption is made that the contribution of the $\bar{K}^{0}$'s to the $K^{0}_S$ yield is negligible. This assumption is justified by the estimate of the antikaon yield performed by the KaoS collaboration in p+Au collisions at the same beam energy of 3.5 GeV: $Y_{K^{-}}/Y_{K^{+}} = 2.3\%$ \cite{Scheinast:2005xs}. Note that this value should be considered as an upper limit for the relative yield of antikaons for both p+p and p+Nb reactions, where in-medium effects, possibly enhancing the antikaon production, are either absent (p+p) or less pronounced (p+Nb) in comparison to p+Au reactions.

\begin{figure}[htb]
\includegraphics[height=0.35\textwidth]{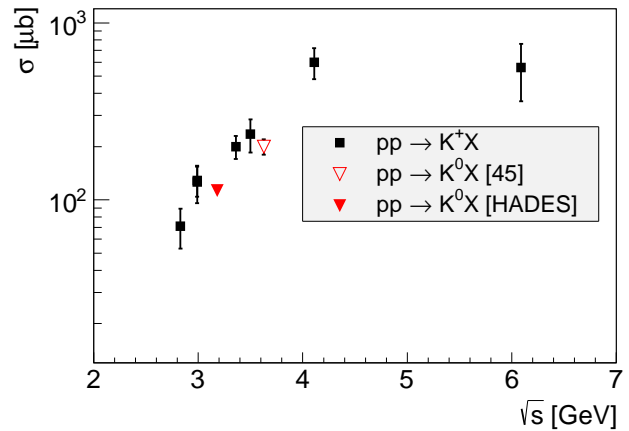}
\caption{\label{fig:pp_cs_tot} (Color online) Energy dependence of the total cross section of inclusive kaon production (squares -- $K^{+}$ \cite{Reed:1968zza}, triangles -- $K^{0}$ \cite{Eisner:1976np}) in proton-proton collisions.} 
\end{figure}

The excitation function of the inclusive kaon production cross section in proton-proton collisions is shown in Fig.~\ref{fig:pp_cs_tot}. The HADES measurement fits well into the trend set by the world data, confirming the validity of the analysis procedure up to the absolute normalization. We note that the available measurements of the $K^{0}_S$ yield at higher energies ($\sqrt{s} > 3.7$~GeV, not shown here) do not allow to deduce the $K^{0}$ production cross section unambiguously, since the $\bar{K}^{0}$ contribution starts to be significant. 

At $\sqrt{s} = 3.18$ in proton-proton collisions (pure isospin-one initial state: $I = 1$, $I_{z} = +1$) the extracted $K^{0}$ ($I_{z} = - 1/2$) yield is $30\%$ lower than the $K^{+}$ ($I_{z} = + 1/2$) yield estimated by an interpolation of two neighbouring measurements.

\section{proton-niobium reactions \label{sec:pNb}}
\subsection{Kaon phase space \label{subsec:pNb_pt}}

\begin{figure*}[t]
\includegraphics[width=0.85\textwidth]{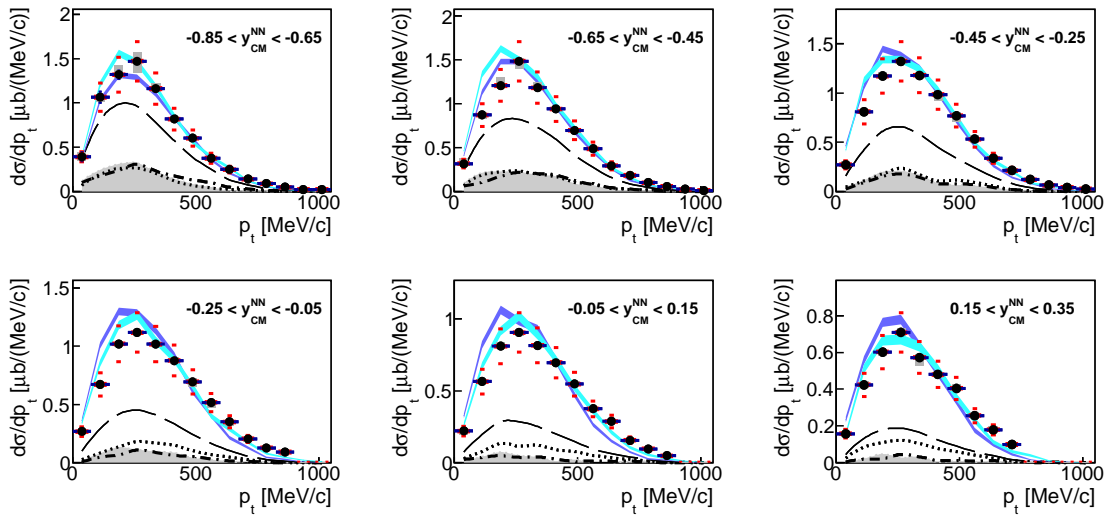}
\caption{\label{fig:pNb_pt_gibuu} (Color online) $K^{0}_S$ transverse momentum spectra in p+Nb collisions: experimental data (black circles) and GiBUU transport model simulations with (cyan) and without (blue) in-medium ChPT KN potential. The widths of the bands indicate the statistical uncertainties of the simulated data sample. The long-dashed curve shows the total contribution of all $K^{0}$ production channels excluding $pp$ and $np$ collisions. Major secondary processes are: $\Delta N$- (dotted curve), $\pi N$-reactions (hatched area) and the contribution from the charge-exchange reactions $K^{+}N \to K^{0}N(\pi)$ (dash-dotted curve).}
\end{figure*}

Transverse momentum spectra of $K^{0}$'s produced in p+Nb collisions are shown in Fig.~\ref{fig:pNb_pt_gibuu} along with GiBUU simulations (with and without repulsive kaon potential, discussed below). 

For the treatment of the $K^{+}$ production and the neutron-proton channel in simulations, the isospin interrelations between different reaction channels as given by the resonance model~\cite{Tsushima:1998jz} are used. It should be stressed that the experimental data on kaon production in neutron-proton collisions are extremely scarce; reported measurements were done with a non-monoenergetic neutron beam \cite{Ansorge:1974ez}. Additionally to the changes in the model described above, we adopt a scaling factor of 0.5 for all three-body processes in the neutron-proton channel, $ n + p \to N + Y + K$. (Further we vary the strengths of these channels before making any interpretation of experimental results.) The resonance model, modified in this way, gives a good description of the experimental data. Contributions of various secondary processes, as implemented in the GiBUU model, are shown in Fig.~\ref{fig:pNb_pt_gibuu} as well.

We note that the spectra shown in Fig.~\ref{fig:pNb_pt_gibuu} might be reasonably approximated by a Boltzmann fit with reduced $\chi^{2}$ values varying from 3 to 5 in different rapidity bins. The extracted slope parameter $T_B(y)$ amounts to 85~MeV at backward rapidity ($y_{CM} \approx -0.8$) and exhibits a maximum of 100 MeV at $y_{CM} \approx -0.2$.

The GiBUU simulations incorporate the repulsive kaon potential resulting from the Chiral Perturbation Theory (ChPT) \cite{Kaplan:1986yq, Nelson:1987dg, Politzer:1991ev, Brown:1992ib, Li:1995as, Schaffner:1996kv}. The ChPT 
model is governed by the $\Sigma_{\text{KN}}$ term appearing in the scalar sector of the in-medium kaon 
interaction and by the pion decay constant $f_{\pi}$ entering into the vector part. 
A range of $\Sigma_{\text{KN}} = 450 \pm 30$~MeV is given in~\cite{Brown:1995qt}; we use a value of 450~MeV. For the 
pion decay constant a reduced in-medium value $f^{*}_{\pi} = \sqrt{0.6}f_{\pi}$ is adopted \cite{Fuchs:2005zg} according to studies of Brown and Rho~\cite{Brown:1995qt}. 
Note that such an in-medium reduction of the pion decay constant is supported from precision spectroscopy of pionic atoms~\cite{Suzuki:2002ae}. The ChPT 
in-medium kaon potential shows a non-linear density dependence, resulting from the corresponding density 
dependence of the effective kaon mass~\cite{Schaffner:1996kv, Prassa:2007zw, Prassa:2009xi}. Note that the ChPT in-medium kaon potential features an explicit momentum dependence. It differs, therefore, from the customary linear parameterization of the potential in terms of the kaon in-medium mass $m^{*}_K = m^{0}_K \left( 1 - \alpha \times \rho_{B}/\rho_{0} \right)$, where $\alpha$ is a parameter (negative for kaons) that governs the strength of the potential. The latter parameterization was used by the IQMD \cite{Hartnack:2010be} and HSD \cite{Cassing:1999es} transport models for the interpretation of heavy-ion \cite{Agakishiev:2010zw} and pion-induced data \cite{Benabderrahmane:2008qs}, respectively. A non-linear density and momentum 
dependence of the in-medium kaon interaction is obtained also by other approaches. Indeed, 
within a One-Boson-Exchange formulation and using the relativistic mean-field approximation, the 
in-medium kaon energy slightly grows with baryon density, in a similar way as the ChPT results but with 
a different curvature~\cite{Schaffner:1996kv}. At normal nuclear density and for the kaon at rest, the ChPT potential results in a magnitude of $\approx 35$~MeV, set by the numerical values of the parameters $\Sigma_{\text{KN}}$ and $f^{*}_{\pi}$.

\begin{figure}[ht]
\includegraphics[width=0.45\textwidth]{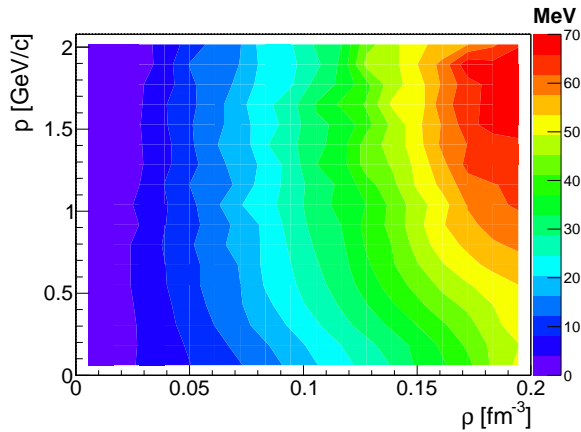}
\caption{\label{fig:U_ChPT_rho_p} (Color online) In-medium ChPT kaon potential $U = E^{*} - E$ (in MeV) as a function of the baryonic density and the kaon momentum.}
\end{figure}

Figure~\ref{fig:U_ChPT_rho_p} illustrates the deviation of the kaon in-medium energy $E^{*}$ from the vacuum energy given by a standard dispersion relation $E = \sqrt{p^{2} + m^{2}}$ as a function of the baryonic density $\rho$ and the kaon momentum $p$. This figure results from GiBUU calculations and shows the approximate region of baryonic densities and momenta probed by kaons in pNb reactions at 3.5~GeV. We note that the potential is significant already in a dilute systems ($\rho \sim \rho_{0}/2$). At higher densities ($\rho \sim \rho_{0}$) a strong momentum dependence resulting from the functional form of the in-medium kaon dispersion relation \cite{Fuchs:2005zg} is visible.

The in-medium potential leads to a rapidity-dependent modification of the simulated $p_{t}$ spectra (Fig.~\ref{fig:pNb_pt_gibuu}). According to the GiBUU model, the apparent effect of the repulsive potential felt by kaons inside the nuclear environment is moderate, which can be attributed to the high beam energy, far above the kaon production threshold, used in this experiment. 

In order to quantify the agreement between the experimental data and the simulations including or excluding the in-medium kaon potential, we perform a $\chi^{2}$-analysis. A covariance matrix is employed where the diagonal entries are statistical and systematic uncertainties added in quadrature and the off-diagonal entries are governed by the global multiplicative normalization uncertainty of 15\%. 

\begin{figure}[htb]
\includegraphics[width=0.5\textwidth]{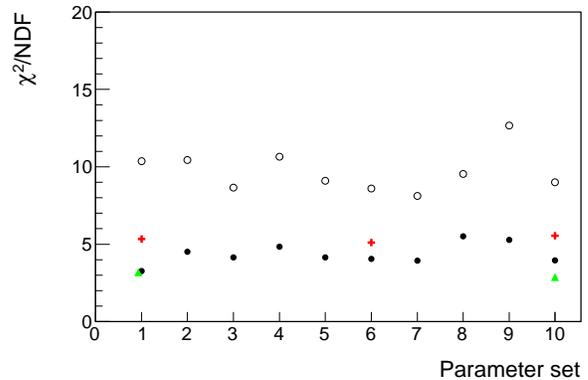}
\caption{\label{fig:pNb_pt_chi2} (Color online) $\chi^{2}$ values for different variations of the parameters entering the model. Empty circles --- simulations without potential, filled circles --- simulations with potential ($\approx 35$~MeV at $\rho_{B} = \rho_{0}$). Crosses correspond to the less repulsive potential ($\approx 25$~MeV) and triangles to the more repulsive one ($\approx 45$~MeV). See text and Table~\ref{table:pNb_pt_chi2_var} for the description of different parameter sets.}
\end{figure}

The $\chi^{2}$ values obtained for the simulations with (filled circles) and without (empty circles) potential are shown in Fig.~\ref{fig:pNb_pt_chi2} as a function of the parameter set number. A significantly lower $\chi^{2}$ value is achieved by the simulations including the in-medium ChPT potential. Furthermore, the strength of the potential has been varied by the choice of the pion decay constant that governs the repulsive vector part of the potential. The $\chi^{2}$ values obtained with a less repulsive version of the potential ($\approx 25$~MeV for the kaon at rest and at normal nuclear density) are indicated by crosses, and those obtained with a more repulsive version ($\approx 45$~MeV) --- by triangles. The data systematically disfavour the weaker potential of $\approx 25$~MeV, whereas the more repulsive potential can not be excluded.

The model includes a number of poorly constrained parameters (mostly kaon production cross sections for the channels that are hard or impossible to measure, such as $np \to NYK$, $\Delta N \to NYK$, etc.). Therefore, it is necessary to study the stability of the results by varying the parameters of the model.  

We performed systematic checks, results of which are shown in Fig.~\ref{fig:pNb_pt_chi2}. The parameter set 1 corresponds to the standard choice of all parameters, as explained above. All other variations are described in Table~\ref{table:pNb_pt_chi2_var}. Each row in this table corresponds to a 25\% variation of a particular channel's strength with respect to the standard values. An exception is the parameter set 10, for which strengths of two channels were varied simultaneously.

\begin{table}[h]%The best place to locate the table environment is directly after its first reference in text
\caption{\label{table:pNb_pt_chi2_var}%
Variations of the model parameters}
\begin{ruledtabular}
\begin{tabular}{lll}
Set & Channel & Variation, \% \\
\colrule
2 & $\sigma(\Delta N \to K X)$ & +25\\
3 & $\sigma(\Delta N \to K X)$ & $-$25\\
4 & $\sigma(\pi N \to K X)$ & +25\\
5 & $\sigma(\pi N \to K X)$ & $-$25\\
6 & $\sigma(n p \to N Y K)$ & +25\\
7 & $\sigma(n p \to N Y K)$ & $-$25\\
8 & $\sigma(K N \to K X)$ & +25\\
9 & $\sigma(K N \to K X)$ & $-$25\\
10 & $\sigma(n p \to N Y K)$ \& & +30 \\
 & $\sigma(NN \to \Delta(1232) Y^{*} K)$ & $-$30\\
\end{tabular}
\end{ruledtabular}
\end{table}

As follows from Fig.~\ref{fig:pNb_pt_chi2}, for all these variations of the input parameters, simulations with the in-medium potential constantly deliver lower $\chi^{2}$ values than simulations without the potential.

\begin{figure}[htb]
\includegraphics[width=0.5\textwidth]{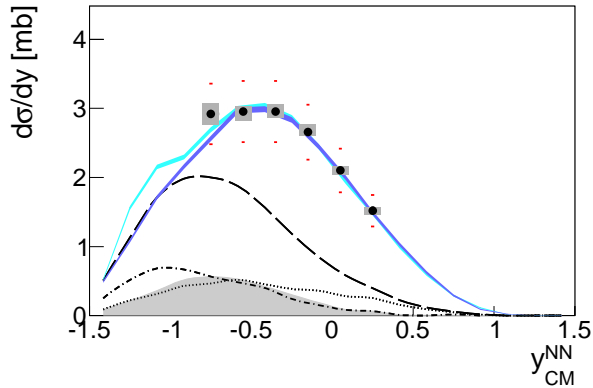}

\caption{\label{fig:pNb_y_gibuu} (Color online) $K^{0}_{S}$ rapidity distribution in the nucleon-nucleon center-of-mass reference frame for $p$+Nb collisions (filled circles) and GiBUU transport model simulations with (cyan) and without (blue) the in-medium ChPT KN potential. The long-dashed curve shows the total contribution of all $K^{0}$ production channels excluding $pp$ and $np$ collisions. Major secondary processes are: $\Delta N$- (dotted curve), $\pi N$-reactions (hatched area) and the contribution from the charge-exchange reactions $K^{+}N \to K^{0}N(\pi)$ (dash-dotted curve).}
\end{figure}

The rapidity distribution of kaons detected in p+Nb collisions is shown in Fig.~\ref{fig:pNb_y_gibuu}. Contrary to the symmetric bell-shaped spectrum in proton-proton reactions, we observe a strong shift of the distribution towards target rapidity. Integration of the simulated distribution, justified by the fact that the experimental data are described well, allows to estimate the inclusive production cross section of neutral kaons:
\begin{equation}
\sigma(p + \text{Nb} \to K^{0}+X ) = 8.3 \pm 1.2~\text{mb},
\end{equation}
where the quoted error represent the dominating absolute normalization uncertainty.

The rapidity distribution is well described by the GiBUU simulations. According to the model, the shape of the distribution is strongly influenced by the re-scattering of the kaons on target nucleons; the rapidity distribution is, therefore, sensitive to the kaon-nucleon scattering cross section. 

We note that the usage of the vacuum KN scattering cross sections allows for a good description of the rapidity distribution. Indeed, as follows from Fig.~\ref{fig:pNb_pt_chi2}, a 25\% variation of these parameters (parameter sets 8 and 9) does not improve the description of the kaon phase space.

The contribution from the charged kaons involved in charge-exchange reactions $K^{+}n \to K^{0}p$, $K^{+}N \to K^{0}N\pi$ is moderate (10\% percent of the observed yield); the model is, therefore, not required to describe the charged kaon yield with a high accuracy.

The influence of the KN in-medium potential on the rapidity spectrum, at least in the experimentally accessible kinematical region, is negligible.

\subsection{Comparison of kaon emission in different colliding systems}
A direct comparison of the p+Nb and p+p measurements has been done for the momentum spectra (in the laboratory reference frame) employing the nuclear modification factor $R_{\text{Nb}/\text{p}}(p)$, defined as
\begin{equation}
R_{\text{Nb}/\text{p}}(p) = \frac{d\sigma^{K^{0}}_{pNb}/dp}{d\sigma^{K^{0}}_{pp}/dp} \times \frac{N^{pp}_{part}}{N^{pNb}_{part}} \times \frac{\sigma^{pp}_{tot}}{\sigma^{pNb}_{tot}},
\end{equation} 
where $d\sigma/dp$ is the differential cross section of $K^{0}$ production in p+Nb and p+p collisions, $N_{part}$ is the number of participants ($N^{pp}_{part} = 2$, $N^{pNb}_{part} = 2.5$) and $\sigma_{tot}$ is the total reaction cross section ($\sigma^{pp}_{tot} = 43.3$~mb \cite{Baldini:1988ti}, $\sigma^{pNb}_{tot} = 848$~mb \cite{Tlusty2010}).

The ratio in terms of $R_{\text{Nb}/\text{p}}(p)$ has already been used in the comparative analysis of the electron-positron pair production in p+p and p+Nb collisions \cite{Agakishiev:2012vj}. There, it was shown that for identified $\omega$ mesons the $R_{\text{Nb}/\text{p}}(p)$ is below unity for all momenta; this observation was interpreted as a hint for absorption of $\omega$ mesons in the nucleus. Since there is no conventional mechanism for kaon absorption in nucleonic environment, the $R_{\text{Nb}/\text{p}}(p)$ for kaons is expected to be mostly sensitive to production mechanisms, including secondary reactions (mostly $\pi N$ and $\Delta N$ collisions), and scattering of kaons.

\begin{figure}[htb]
\includegraphics[width=0.45\textwidth]{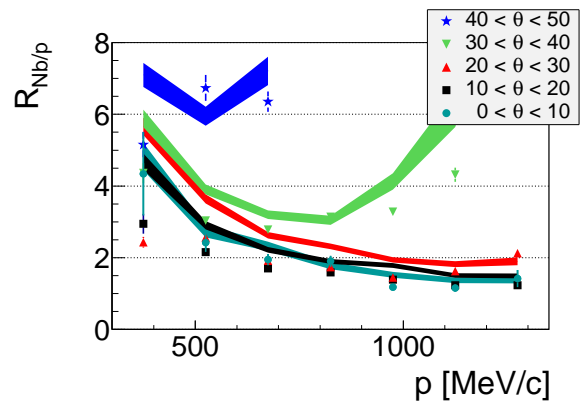}
\caption{\label{fig:pNb_pp} (Color online) Momentum dependence of the nuclear modification factor $R_{\text{Nb}/\text{p}}(p) \propto \sigma^{K^{0}}_{p\text{Nb}}/\sigma^{K^{0}}_{pp}$. Symbols correspond to the experimental data and bands to GiBUU simulations. Only statistical uncertainties are shown.}
\end{figure}

The nuclear modification factor obtained for $K^{0}$ meson is presented in Fig.~\ref{fig:pNb_pp} for five bins of the kaon polar angle in the laboratory reference frame ($R_{\text{Nb}/\text{p}}(p) = f \left(p, \theta \right)$) along with the corresponding GiBUU simulations. At forward angles ($0^{\circ}<\theta<30^{\circ}$) in the region $p_{K^{0}}>800$~MeV$/c$ a fall of the kaon yield in p+Nb collisions, caused by the scattering on nucleons, is visible. The rapid rise of the nuclear modification factor at larger angles ($30^{\circ}<\theta<40^{\circ}$) is explained by the Fermi motion and other nuclear effects (such as neutron-proton collisions and secondary reactions), which allow production of more energetic kaons in p+Nb collisions relative to the p+p case. 

A comparison of the HADES p+Nb results with the data on $K^{+}$ production in p+$^{197}$Au and p+$^{12}$C collisions reported by the KaoS collaboration at the same beam kinetic energy of 3.5 GeV~\cite{Scheinast:2005xs} was made in terms of $R_{\text{Au}/\text{Nb}}(p)$ and $R_{\text{Nb}/\text{C}}(p)$ (defined in full analogy to $R_{\text{Nb}/\text{p}}(p)$). All relevant quantities are listed in Table~\ref{table:Rpa}.

\begin{table}[b]%The best place to locate the table environment is directly after its first reference in text
\caption{\label{table:Rpa}%
Information about kaon measurements in proton-proton and proton-nucleus collisions ($E^{p}_{kin}=3.5$~GeV) performed by the KaoS~\cite{Scheinast:2005xs} and the HADES collaborations. The number of participants is estimated by a geometrical overlap model \cite{Eskola:1988yh}; the total reaction cross sections for the p+Au and p+C collisions are estimates using the parameterization from \cite{Tripathi:1996jn}.
}
\begin{ruledtabular}
\begin{tabular}{llll}
Experiment/Particle & System & $N^{min. bias}_{part}$ & $\sigma_{tot}$, mb\\
\colrule
KaoS/$K^{+}$ & p+$^{197}$Au & 3.1 & 1616\\
KaoS/$K^{+}$ & p+$^{12}$C & 2.1 & 243.4\\
HADES/$K^{0}$ & p+$^{93}$Nb & 2.5 & 848\\
HADES/$K^{0}$ & p+p & 2 & 43.3\\
\end{tabular}
\end{ruledtabular}
\end{table}

\begin{figure}[htb]
\includegraphics[width=0.45\textwidth]{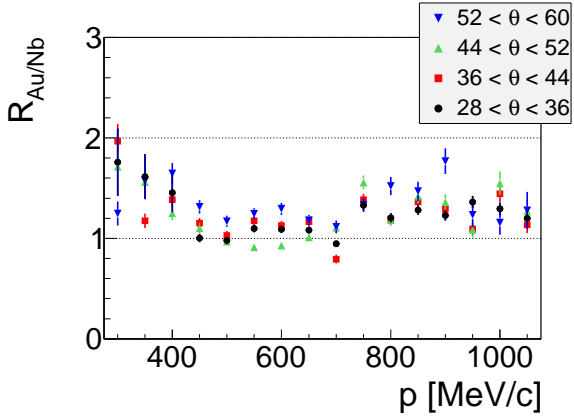}
\caption{\label{fig:RpA_pAu_pNb} (Color online) Nuclear modification factor $R_{\text{Au/Nb}} \propto \sigma^{K^{+}}_{p\text{Au}}/\sigma^{K^{0}}_{p\text{Nb}}$. Only statistical uncertainties (originating from both HADES and KaoS measurements) are shown.}
\end{figure}

Figure~\ref{fig:RpA_pAu_pNb} shows a comparison of $K^{+}$ momentum spectra measured in p+Au collisions (KaoS data) with $K^{0}$ spectra in p+Nb (HADES data). For all angular bins the nuclear modification factor shows the same trend: it is close to unity, which is a good benchmark for the analysis and normalization procedures.

\begin{figure}[t]
\includegraphics[width=0.45\textwidth]{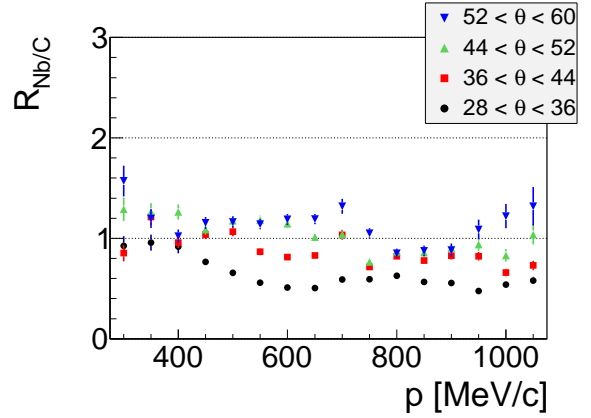}
\caption{\label{fig:RpA_pNb_pC} (Color online) Nuclear modification factor $R_{\text{Nb/C}}(p) \propto \sigma^{K^{0}}_{p\text{Nb}}/\sigma^{K^{+}}_{p\text{C}}$.}

\end{figure}

The comparison between p+Nb ($K^{0}$'s) and p+C data ($K^{+}$'s) is shown in Fig.~\ref{fig:RpA_pNb_pC}. In this case a systematic behavior of $R_{\text{Nb}/\text{C}}(p)$ is observed---it grows with an increase of the polar angle. This effect originates mostly from neutron-proton reactions, more probable on niobium than on carbon target, reinforced by the scattering of forward kaons to larger polar angles. 
A large ratio of the radii of the $^{93}$Nb and $^{12}$C targets ($r_{\text{Nb}}/r_{\text{C}}\sim(93/12)^{1/3}\approx2$) explains why this effect is more pronounced than in the Au/Nb case ($r_{\text{Au}}/r_{\text{Nb}}\sim(197/93)^{1/3}\approx1.3$).


\subsection{Effect of the potential on the ratio of momentum spectra}

As shown above the action of the potential is visible in the $p_{t}$-$y$ phase space. To support this statement we consider another (but not independent) observable. This is the ratio $\mathcal{R}$ of $K^{0}$ momentum spectra measured in p+Nb collisions in two adjacent bins of the polar angle (both the momentum and the polar angle are in the laboratory reference frame): 

\begin{equation}
\mathcal{R} = \frac{d\sigma/dp\left(10^{\circ} < \theta < 20^{\circ}\right)}{d\sigma/dp\left(20^{\circ} < \theta < 30^{\circ}\right)}.
\end{equation}

\begin{figure}[h]
\includegraphics[width=0.5\textwidth]{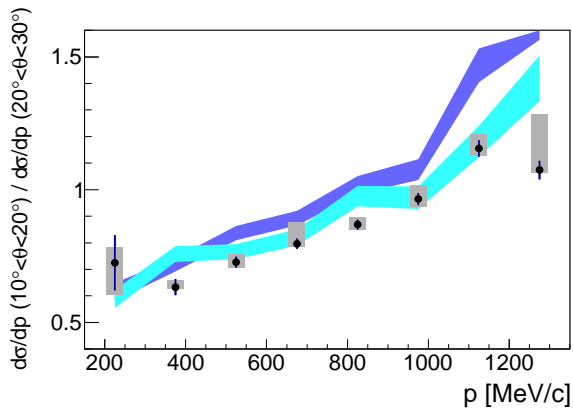}
\caption{\label{fig:pNb_ratio_pot_gibuu} (Color online) Ratio of $K^{0}$ momentum spectra reconstructed in two bins of the polar angle (black circles) and GiBUU transport model simulations with (cyan) and without (blue) repulsive kaon potential.}
\end{figure}

It is shown in Fig.~\ref{fig:pNb_ratio_pot_gibuu} along with GiBUU simulations. According to the GiBUU model, the effect of the repulsive potential is maximal for the forward kaons (which travel, on average, a longer path in the nucleus), whereas for the kaons emitted at larger polar angles ($\theta > 20^{\circ}$) the influence of the potential is much smaller. This observation explains the choice of the polar angle bins used for the ratio. An important feature of this ratio is that it is by construction free from the systematic uncertainty originating from the absolute normalization. Moreover, it is expected to be less sensitive to the ambiguities in the kaon production and interaction cross sections implemented in the transport model. The GiBUU simulation including ChPT kaon potential gives a better description of the experimental data in this representation as well. 

\begin{figure}[h]
\includegraphics[width=0.5\textwidth]{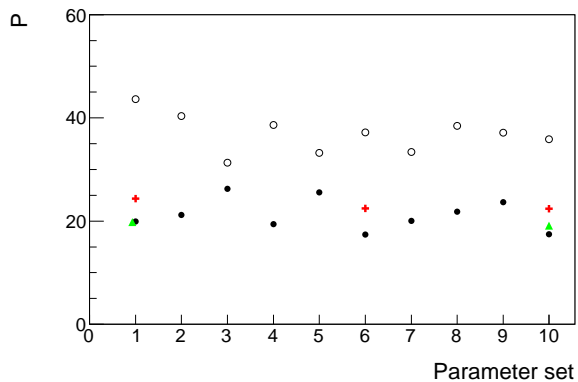}
\caption{\label{fig:pNb_p_ratio_pull} (Color online) Pull values (defined by (Eq.~\ref{eq:pull})) for different variations of the parameters entering the model. Empty circles --- simulations without potential, filled circles --- simulations with potential ($\approx 35$~MeV at $\rho_{B} = \rho_{0}$). Crosses correspond to the less repulsive potential ($\approx 25$~MeV) and triangles to the more repulsive one ($\approx 45$~MeV). See text and Table~\ref{table:pNb_pt_chi2_var} for the description of different parameter sets.}
\end{figure}

We repeat the very same variations of the model parameters as for the $p_{t}$-$y$ data (cf. Fig.~\ref{fig:pNb_pt_chi2}). In this case, however, due to the strong asymmetry of the systematic uncertainties, we employ a pull value $P$ defined as:

\begin{equation}
\label{eq:pull}
P = \sum_{i} \frac {| r_{i}^{\text{exp}} - r_i^{\text{sim}} |} {\sigma_{i}^{\pm}},
\end{equation}
where the upper experimental uncertainty $\sigma_{i}^{+}$ is taken if the simulated data point $r_{i}^{\text{sim}}$ lies above the experimental one, and the lower uncertainty $\sigma^{-}$ is taken in the opposite case. The results of this analysis are summarized in Fig.~\ref{fig:pNb_p_ratio_pull}. Again (cf. Fig.~\ref{fig:pNb_pt_chi2}), the simulations with the in-medium potential always provide lower pull values as compared to the no-potential case, confirming the observation of the potential effect. The pull values, however, exhibit a larger scattering of statistical origin. Finally, the weaker potential ($\approx 25$~MeV) delivers worse pull values in comparison to the standard one used in this analysis ($\approx 35$~MeV), whereas the stronger potential ($\approx 45$~MeV) describes the data almost equally well.

\section{Summary and conclusions \label{sec:sum}}

The experimental results obtained from p+Nb collisions at $E_{beam}=3.5$~GeV, both the phase-space distributions and the ratio of momentum spectra, strongly support the ChPT prediction of a repulsive potential, the effect of which is evaluated by the GiBUU transport model. For the kaon at rest and at normal nuclear density, the ChPT potential amounts to $\approx 35$~MeV, consistent with the result reported by the HADES collaboration in the analysis of heavy-ion reactions \cite{Agakishiev:2010zw}. This statement remains stable under systematic variations of those model parameters that are not known precisely. Our analysis, furthermore, demonstrates the importance of the momentum dependence of the in-medium potential on the treatment of fast kaons. 
 
In order to validate the simulation model, the results on the inclusive production of $K^0$ mesons in proton-proton collisions were examined in a broad region of the phase space. It was shown that the resonance model for kaon production in baryon-baryon collisions \cite{Tsushima:1998jz} significantly overestimates the inclusive production of $K^{0}$ mesons in proton-proton collisions. The reason is that the model overestimates cross sections for a number of individual channels, for example the reaction $p + p \to p + \pi^{+} + \Lambda + K^{0}$. The new data demand that the $K^{0}$ production channels with 5-body final states must be included at this energy. 
The corresponding cross sections have been tuned in the GiBUU transport model such as to reproduce the experimental p+p data. 
The GiBUU simulations with the modified model are able to reproduce the data obtained in proton-niobium collisions. 

A strong shift of the rapidity distribution towards target rapidity was observed in proton-niobium reactions. According to the GiBUU model, this shift is mainly caused by the kaon-nucleon scattering together with secondary $\pi N$ and $\Delta N$ reactions. The GiBUU simulations employing vacuum KN scattering cross sections describe the rapidity distribution very well.

The developed simulation model implemented in the GiBUU code can be used for a further, refined analysis of neutral kaon production in heavy ion collisions characterized by high baryonic densities.

\begin{acknowledgments}
The HADES collaboration gratefully acknowledges the support by the grants LIP Coimbra, Coimbra (Portugal) PTDC/FIS/113339/2009, SIP JUC Cracow, Cracow6 (Poland): N N202 286038 28-JAN-2010 NN202198639 01-OCT-2010, HZ Dresden-Rossendorf (HZDR), Dresden (Germany) BMBF 06DR9059D, TU M\"unchen, Garching (Germany) MLL M\"unchen: DFG EClust 153, VH-NG-330 BMBF 06MT9156 TP5 GSI TMKrue 1012 NPI AS CR, Rez, Rez (Czech Republic) MSMT LC07050 GAASCR IAA100480803, USC - S. de Compostela, Santiago de Compostela (Spain) CPAN:CSD2007-00042, Goethe University, Frankfurt (Germany): HA216/EMMI HIC for FAIR (LOEWE) BMBF:06FY9100I GSI F\&E EU Contract No. HP3-283286.
\end{acknowledgments}

\bibliography{K0_pp_pNb}

\end{document}